\documentclass[fleqn,10pt]{wlscirep}
\usepackage[LGRgreek]{mathastext}

\newcommand{\fig}[1]{Fig.~\ref{#1}}
\newcommand{\be}[1]{\begin{equation}\label{#1}}
\newcommand{\ee}{\end{equation}}

\title{Non-sequential double ionization with near-single cycle laser pulses}

\author[1]{A. Chen}
\author[2,4]{M. K\"{u}bel}
\author[3,4]{B. Bergues}
\author[3,4]{M. F. Kling}
\author[1,*]{A. Emmanouilidou}
\affil[1]{Department of Physics and Astronomy, University College London, Gower Street, London WC1E 6BT, United Kingdom}
\affil[2]{Joint Laboratory for Attosecond Science, University of Ottawa and National Research Council, 100 Sussex Drive, Ottawa, Ontario, Canada K1A 0R6 }
\affil[3]{Max-Planck-Institut f\"{u}r Quantenoptik, D-85748 Garching, Germany}
\affil[4]{Department f\"{u}r Physik, Ludwig-Maximilians-Universit\"{a}t, D-85748 Garching}

\affil[*]{a.emmanouilidou@ucl.ac.uk}

\begin{abstract}
A three-dimensional semiclassical model is used to study double ionization of Ar when driven by a near-infrared and near-single-cycle laser pulse  for intensities ranging from 0.85$\times$10$^{14}$ W/cm$^{2}$ to 5$\times$10$^{14}$ W/cm$^{2}$. Asymmetry parameters, distributions of the sum of the two electron momentum components 
 along the direction of the polarization of the laser field  and correlated momenta are computed as a function of intensity and of the carrier envelope phase. A very good agreement is found with recently obtained results in kinematically complete experiments employing near-single-cycle laser pulses. Moreover, the contribution of  the direct and   delayed pathways of double ionization is investigated for the above observables. Finally, an experimentally obtained  anti-correlation momentum pattern at higher intensities is reproduced with the three-dimensional semiclassical model and shown  to be due to a transition from strong to soft recollisions with increasing intensity.
 \end{abstract}
\begin{document}

\flushbottom
\maketitle
%
%
\thispagestyle{empty}

\section*{Introduction}
Non-sequential double ionization (NSDI) in intense near-infrared laser fields is a fundamental process with  electron-electron correlation playing a key role \cite{Corkum,NSDI1,NSDI2}.  Considerable information regarding NSDI has been obtained from  kinematically complete experiments, i.e.,  the momenta of the escaping  electrons and ions are measured in coincidence  \cite{kinematic1}. Most of these experiments employ multi-cycle laser pulses allowing for multiple recollisions to occur before both electrons ionize. Multiple recollisions complicate the electron dynamics and render the comparison with theory difficult. Recently, however, kinematically complete experiments succeeded in confining  NSDI to a single laser cycle by using carrier-envelope phase (CEP)-controlled few- and near-single-cycle pulses \cite{Camus,Kling1}. These experiments with near-single-cycle pulses allow for an easier comparison between theory and experiment.

To interpret the double ionization spectra of driven Ar measured using near-single-cycle laser pulses,  a simple one-dimensional (1D) classical model was put forth \cite{Kling1,Kling2, Bergues}. This model relies on the assumption  that the dominant pathways of double ionization are, for small and intermediate intensities,  delayed non-sequential ionization and, for higher intensities, sequential ionization. This model neglects the contribution of another major pathway of double ionization, namely,  direct ionization as well as the Coulomb potential.    This 1D model did not achieve  a quantitative agreement with the complete set of available experimental data  over the whole intensity range. Delayed ionization---also referred to as recollision-induced excitation with subsequent field ionization, RESI \cite{RESI1,RESI2}, and direct ionization are  two main pathways of NSDI. An interesting finding of these near-single cycle experiments  was that the  correlated momenta components of the two escaping electrons along the direction of the laser field  have a  cross-shaped pattern for an intensity around 10$^{14}$ W/cm$^2$ \cite{Kling1,Kling2, Bergues}. A cross-shaped correlated momenta pattern due to the delayed double ionization mechanism was previously identified in the context of strongly-driven He at an intensity 9$\times$10$^{14}$ W/cm$^{2}$ and a wavelength 400 nm \cite{Emmanouilidou1}. In the context of strongly-driven Ar,  the above described  1D model attributed the cross-shaped pattern of the correlated momenta  to the delayed pathway of double ionization  \cite{Kling2, Bergues}. Furthermore, a quantum mechanical calculation, which only considers the delayed pathway of double ionization  and neglects  the Coulomb potential, identified the key role that the symmetry of the excited state plays in the final shape of the correlated momenta \cite{Faria}. 

In this work, using  a three-dimensional (3D) semiclassical model, NSDI of Ar is studied  when Ar is driven by  750 nm  near-single-cycle laser pulses at intensities ranging from 0.85$\times$10$^{14}$ W/cm$^2$ to 5$\times$10$^{14}$ W/cm$^2$. In this 3D model the only approximation is in the initial state. There is no approximation during the time propagation. That is,    
all Coulomb forces and the interaction of each electron with the laser field are fully accounted for. No assumptions are made regarding the prevailing mechanism of double ionization and there is no free parameter. This is not the case for  the 1D model 
\cite{Bergues}. Moreover, the Coulomb singularity is fully accounted for using  regularized coordinates \cite{KS}. This is an advantage over  models which soften the Coulomb potential  \cite{Huang}. 
 Previous successes of this 3D model include identifying  the mechanism responsible for the fingerlike structure \cite{Agapi1}, which  was predicted theoretically \cite{Taylor1} and was observed experimentally for He driven by 800 nm laser fields \cite{vshape1,vshape2}. Moreover, this model was used to investigate direct versus delayed pathways of NSDI for He driven by a 400 nm laser field while achieving excellent agreement with fully ab-initio quantum mechanical calculations \cite{Agapi2}. Using this model,  in this work, several observables are computed  for different intensities of strongly-driven Ar. These observables are the sum of the two electron momentum components along the direction of the polarization of the laser field and the double differential probability of the two electron momentum components along the polarization direction of the laser field. Furthermore,  the amplitude 
and the phase of the asymmetry parameter that determines the difference of the ions escaping with positive versus negative momentum along the polarization direction of the laser field are computed as a function of the carrier envelope phase (CEP) and the intensity.  The computed results using the 3D semiclassical model are found to be in better agreement with the experimental results over the whole intensity range  \cite{Kling1,Kling2, Bergues} compared to the results obtained with the 1D model  in ref. \cite{Bergues}.  Motivated by the good agreement between theory and experiment, the strength of the 3D semiclassical model in fully accounting for the  electron dynamics is utilized to identify the prevailing pathway of double ionization as a function of intensity. In addition, for a small intensity around 10$^{14}$ W/cm$^{2}$, the dependence of   the double ionization pathways  on CEP is computed using the 3D semiclassical model. Finally,  the transition from strong to soft recollisions is identified as the main reason for  the experimentally observed escape of the two  electrons  with opposite velocities  at higher intensities \cite{Agapi4}. 

\section{Method}
 For the current  studies,  a 3D semiclassical model is employed that is formulated in the framework of the dipole approximation \cite{Agapi1}. The time propagation is determined by the three-body Hamiltonian of the two electrons with the nucleus kept fixed.   All Coulomb forces are accounted for: the interaction of each electron with the nucleus and the laser field and the electron-electron interaction are all included in the time propagation.  The laser field      is given by

\begin{eqnarray}
\vec{E}(t)=E_0e^{\left(-2ln2\left(\frac{t}{\tau}\right)^2\right)}cos(\omega t+\phi)\hat{z},
\end{eqnarray}
where $\tau=4fs$ is the full-width-half-maximum pulse duration, $\omega$=0.061 a.u (750 nm) the frequency, $E_{0}$ the strength  and $\phi$  the CEP of the laser field.
 In this work  linearly polarized laser fields are considered. 
The initial state in the 3D model  entails one electron tunneling through the field-lowered Coulomb potential with  the Ammosov-Delone-Krainov (ADK) formula \cite{A1,A2}. To obtain the tunnel ionization rate for Ar, in the ADK formula  the first ionization energy of Ar, i.e. $I_{p_{1}}=0.579$ a.u. and the effective charge $Z=1$ are used.   The momentum along the direction of the electric field is zero while the transverse one is given by a Gaussian distribution  \cite{A1,A2}. The remaining electron is initially described by a microcanonical distribution \cite{Abrimes}. The microcanonical distribution is obtained using the second ionization energy of Ar, i.e.  $I_{p_{2}}=1.02$ a.u. and an effective charge equal to $Z=2$ a.u. During the time propagation each electron is interacting with the nucleus with charge $Z=2$.  In what follows,  the tunneling and bound electron are denoted as electrons 1 and 2, respectively.  

The intensities considered for the results presented  range from 0.85$\times$10$^{14}$ W/cm$^2$ to 5$\times$10$^{14}$ W/cm$^2$. For the smallest intensity, 12 CEPs are considered ranging from $\phi=15^{\circ}$ to $\phi=345^{\circ}$ in steps of 30$^{\circ}$. For all other intensities, 24 CEPs are considered ranging from $\phi=0^{\circ}$ to $\phi=360^{\circ}$ in steps of 15$^{\circ}$. For the results  presented regarding total double ionization  the average has been taken over all CEPs for each intensity. It is noted that the computations required are challenging, since, it is time-consuming to obtain enough double ionization events that render the statistical error very small for each of the 12 or 24 CEPs for each intensity. Therefore, computations were performed for six intensities in the range from 0.85$\times$10$^{14}$ W/cm$^2$ to 5$\times$10$^{14}$ W/cm$^2$. Using the results obtained at these six intensities   an average over the focal volume is performed \cite{laser_volume_effect} to  directly compare with experiment. It is, however, noted that computations at a larger number of intensities are needed to  account more accurately for the focal volume effect. For the results presented  it is stated explicitly when focal volume averaging is included and when it is not.

\section{Results}
\subsection{Double ionization and pathways}
 
\begin{figure} [ht!]
\centering
 \includegraphics[clip,width=0.40\textwidth]{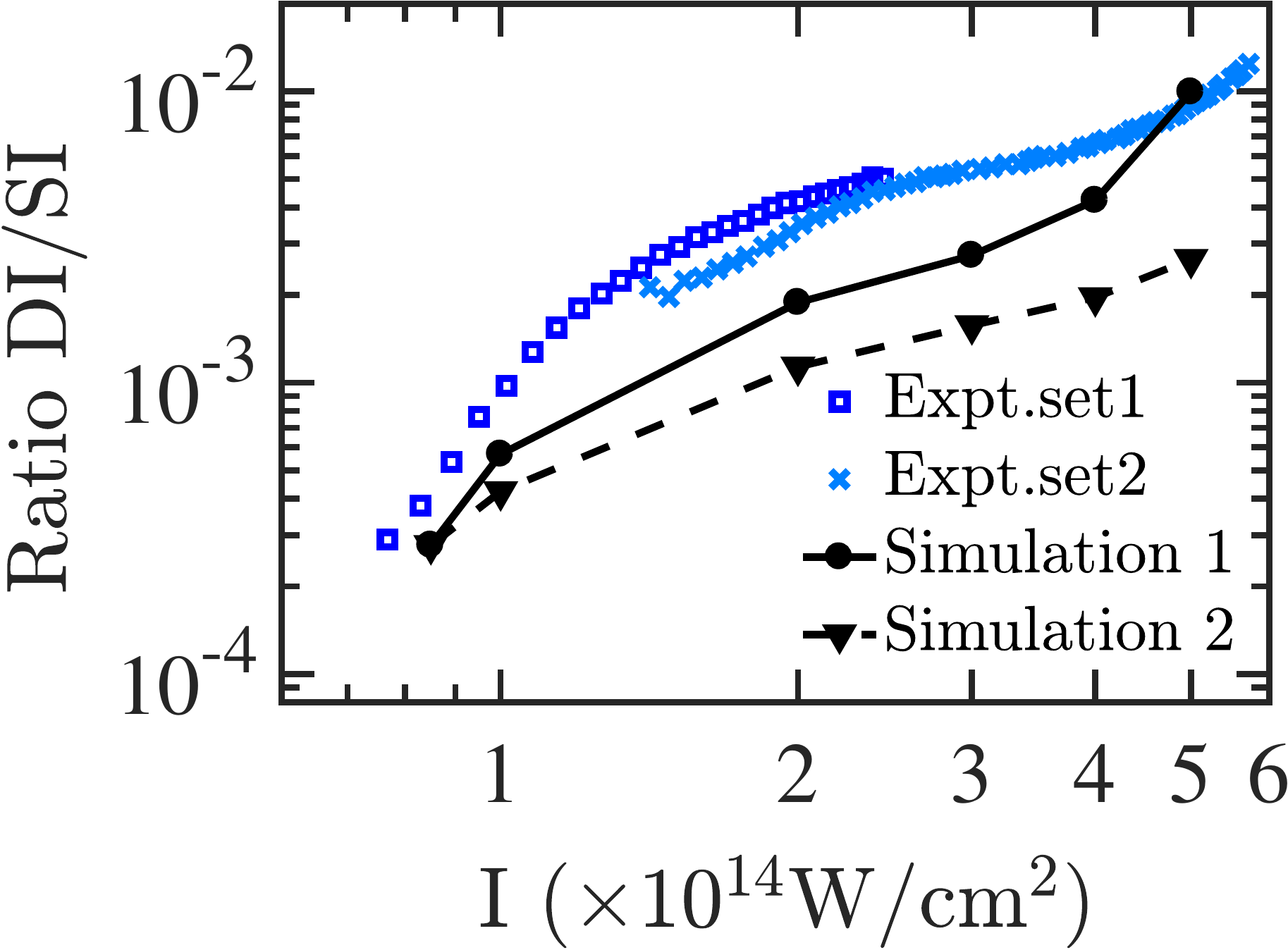}
\caption{Ratio of double to single ionization as a function of intensity. Experimental results are denoted by dark blue squares and light blue crosses and computed results are presented by a solid line with black circles when the focal volume effect is not accounted for and by a dashed-line with triangles when the focal volume effect is accounted for. The difference in the two experimental sets results from slightly different  averaging over the focal volume \cite{Bergues}.}
\label{ratioDISI}
\end{figure}

  In \fig{ratioDISI}, the ratio of double to single ionization events is computed as a function of the laser intensity and compared to the experimental results \cite{Bergues}. It is found that the computed ratio of double to single ionization events reproduces well the overall pattern of the observed ratio. The computed ratio is found to be  at  most a factor of two smaller than the observed ratio and by a factor of 3.5 when the focal volume effect is accounted for.  This difference possibly suggests that the effective charge of $Z=2$ used to model the attractive Coulomb potential in the 3D semiclassical model during time propagation overestimates the Coulomb attraction.
 
\begin{figure}[ht!] 
\centering
 \includegraphics[clip,width=0.40\textwidth]{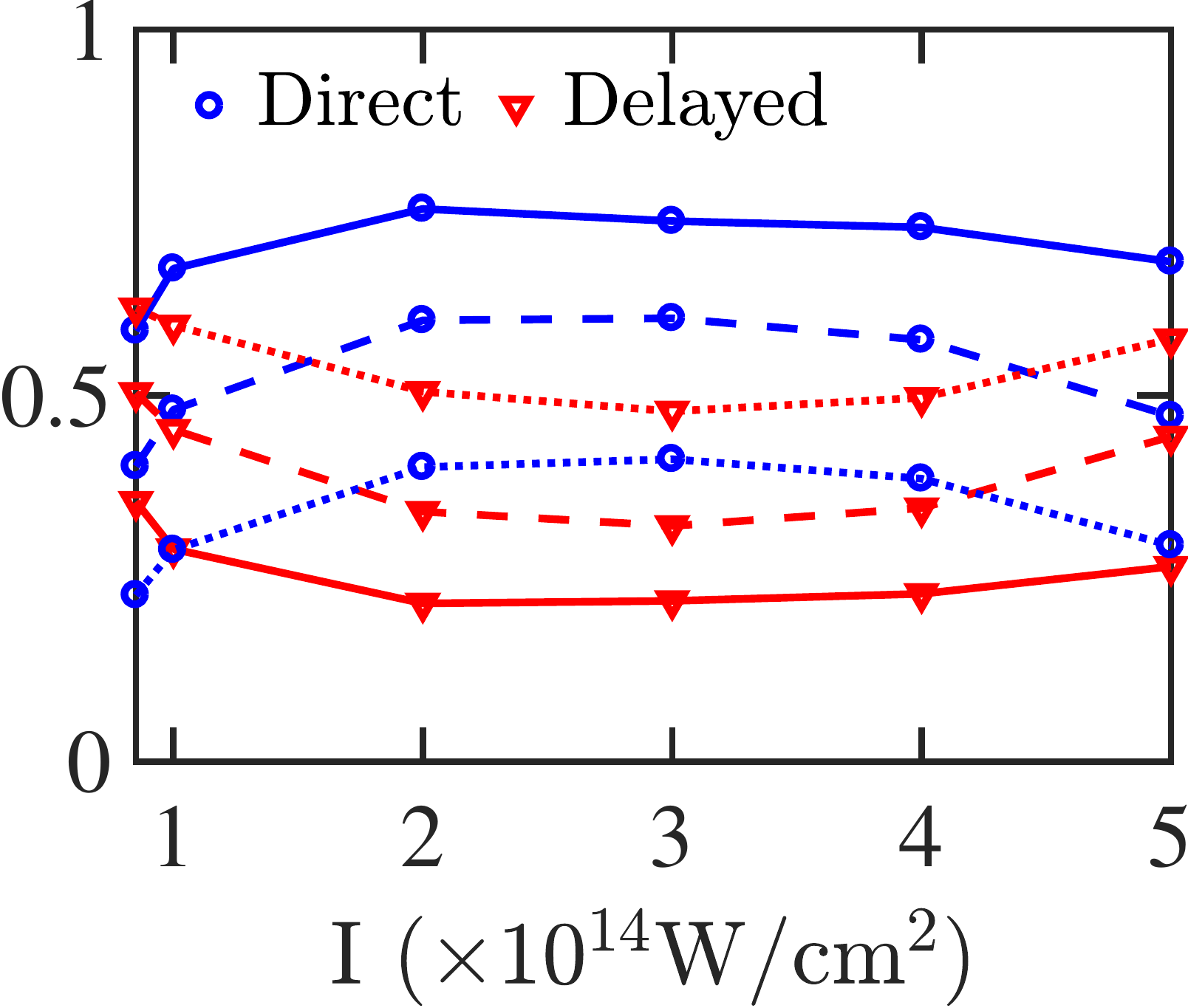}
\caption{Probability of direct (blue circles) and delayed (red triangles) pathways of double ionization as a function of the laser intensity for a delay time of  1/10 T (solid lines), 1/20 T (dashed lines) and 1/40T (dotted lines). The focal volume effect is not accounted for.}
\label{probpathways}
\end{figure}

 Once the doubly ionizing events are obtained using the 3D model, an analysis of the trajectories is performed in time in order to identify
the contribution of the direct and  the delayed pathway of NSDI  as a function of the laser intensity. The main two double ionization energy transfer pathways are identified by using the  time difference between the recollision time $\mathrm{t_{rec}}$ and the ionization time of each electron $\mathrm{t_{ion}^{i}}$, with $\mathrm{i=1,2}$, for each doubly ionizing classical trajectory. The recollision time is defined as the time of minimum approach of the two electrons and is identified by the maximum in the electron pair potential energy. The ionization time for each electron is defined as the time when the sum of the electrons' kinetic energies (using the canonical momentum) and the potential energy due to the electron's interaction with the nucleus becomes positive and remains positive thereafter. The canonical momentum of an electron is given by $\mathrm{{\bf p}-{\bf A}}$, with ${\bf A}$ the vector potential.  The ionization time of the tunneling electron is, thus, not necessarily the time this electron tunnels at the start of the time propagation. This energy  is referred to as compensated energy and was introduced in ref. \cite{Leopold}. A doubly ionized trajectory is labeled as delayed or direct depending on the time differences $\mathrm{t_{ion}^{1}-t_{rec}}$ and $\mathrm{t_{ion}^{2}-t_{rec}}$. Specifically, if the conditions  
\begin{eqnarray}
|t_{ion}^{1}-t_{rec}|<t_{diff}, \hspace{2pt} t_{ion}^{2}<t_{ion}^{1} \hspace{5pt} or \hspace{5pt}  |t_{ion}^{2}-t_{rec}|<t_{diff},  \hspace{2pt}t_{ion}^{1}<t_{ion}^{2}
\end{eqnarray}
are satisfied then the trajectory is labeled as direct. If the conditions 
\begin{eqnarray}
t_{ion}^{1}-t_{rec}>t_{diff}, \hspace{2pt} t_{ion}^{2}-t_{rec}<t_{diff}  \hspace{5pt} or \hspace{5pt} t_{ion}^{2}-t_{rec}>t_{diff}, \hspace{2pt} t_{ion}^{1}-t_{rec}<t_{diff} 
\end{eqnarray}
are satisfied then the trajectory is labeled as delayed. If the conditions 
\begin{eqnarray}
t_{ion}^{1}-t_{rec}>t_{diff}  \hspace{5pt} and \hspace{5pt} t_{ion}^{2}-t_{rec}>t_{diff}
\end{eqnarray}
are satisfied then the trajectory is labeled as double delayed.  The percentage of delayed and direct trajectories  depends on the choice of the time difference $\mathrm{t_{diff}}$. This is shown  in \fig{probpathways} where the probability of direct and delayed double ionization events is plotted for  $\mathrm{t_{diff}}$ equal to 1/10 T, 1/20 T and 1/40 T. The probability of  the direct (delayed) ionization pathway is obtained by dividing the number of doubly ionizing trajectories labeled as direct (delayed) with   the total number of doubly ionizing trajectories. It is found that the percentage  contribution of the direct and the delayed double ionization pathways as a function of intensity display  general trends  that do not significantly depend on the  choice of $\mathrm{t_{diff}}$. Both the  direct and the delayed pathways of double ionization  significantly contribute at all intensities. Thus the direct pathway can not be neglected as was done in previous models. The direct pathway contributes the most for intermediate intensities.  In  \fig{probpathways}, at a high intensity above 4$\times$10$^{14}$ W/cm$^2$, it is shown  that the percentage contribution of the direct pathway of double ionization starts decreasing. At this high intensity a transition from strong to soft recollisions takes place, as discussed in the following.   $\mathrm{t_{diff}}$ can not be chosen neither very large, such as 1/4 T, or very small  such as 1/40 T. Choices in between are reasonable and lead to  similar trends of the   two prevailing pathways of double ionization. $\mathrm{t_{diff}=1/10}$ T is chosen for the results presented in this work. It is found that double delayed trajectories contribute no more than 15\% for the smallest intensity even when the time difference is chosen small and equal to 1/40 T. If instead of the compensated energy the energy of each electron is used to identify the ionization time different results are obtained. 
Namely, one finds that at an intensity of 0.85$\times$10$^{14}$ W/cm$^2$ almost all trajectories are identified as double delayed. This was the conclusion in ref. \cite{Camus}.   Using the actual energy  to identify the ionization time
at an intensity of 3$\times$10$^{14}$ W/cm$^2$ results in the direct pathway of double ionization still only contributing 20\%. However, this is not a reasonable result.  At 3$\times$10$^{14}$ W/cm$^2$ 3.17 U$_{p}$ is equal to 50 eV which is much higher than the second ionization energy of Ar. Moreover, the recollsion at this intensity is strong, which is discussed in {\bf 2.4},  and so the direct pathway of double ionization should contribute significantly. Thus, the compensated energy is employed to identify the ionization time in this work which leads to both the direct and delayed pathway being the main pathways of double ionization in agreement with ref. \cite{Huang} for the smallest intensity.

 \subsection{Distribution of the sum of the momenta}
\begin{figure} [ht!]
\centering
 \includegraphics[clip,width=0.50\textwidth]{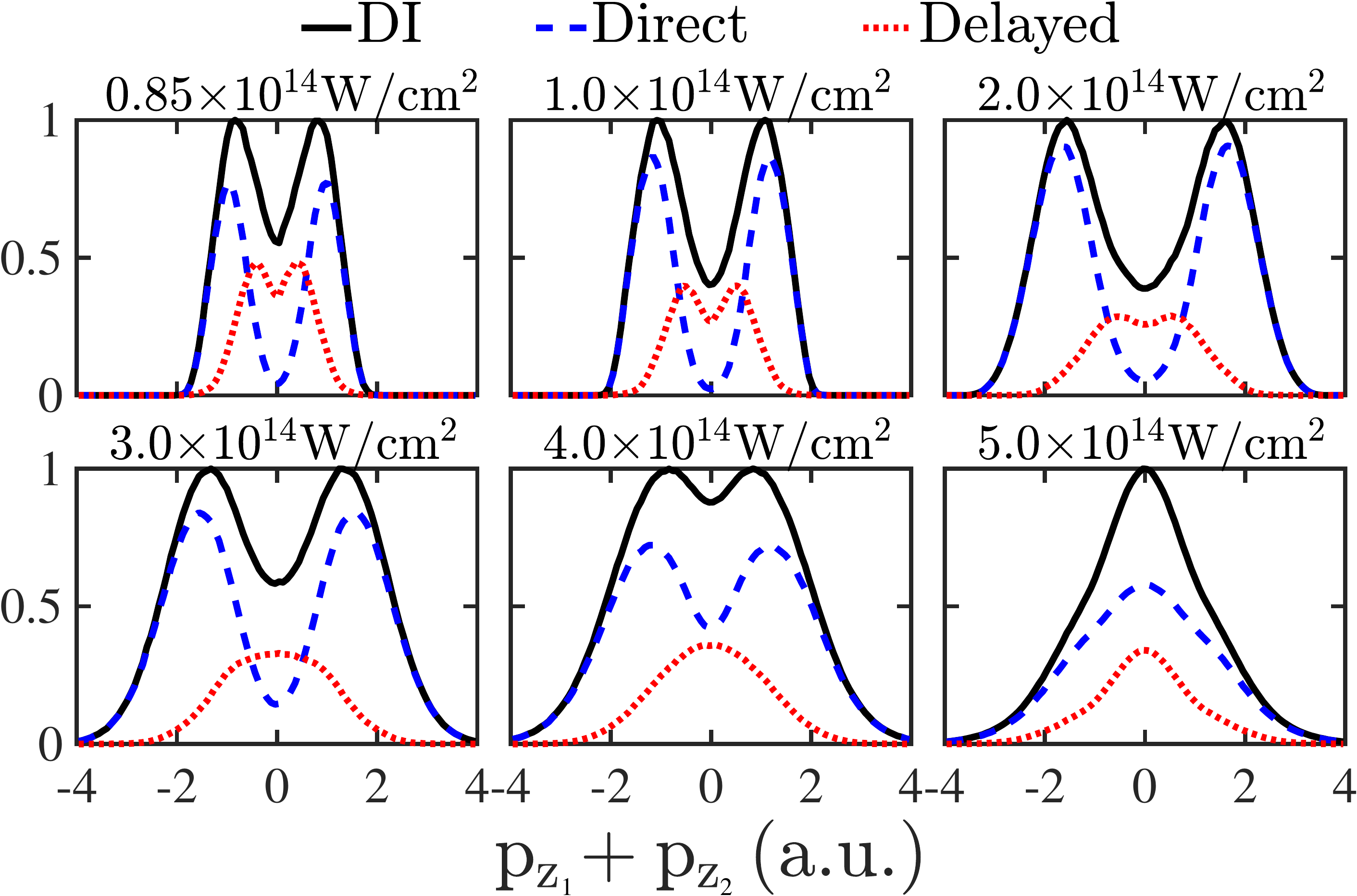}
\caption{Distribution of the sum of the two electron momentum components parallel to the polarization of the laser field (black solid lines) for laser field intensities from 0.85$\times$10$^{14}$ W/cm$^{2}$ to 5$\times$10$^{14}$W/cm$^2$. For each intensity, the distribution of the sum of the momenta of the delayed pathway (red dotted lines) and of the direct pathway (blue dashed lines) are also plotted. The focal volume effect is not accounted for.}
\label{summomenta1}
\end{figure}

\begin{figure} [ht!]
\centering
 \includegraphics[clip,width=0.50\textwidth]{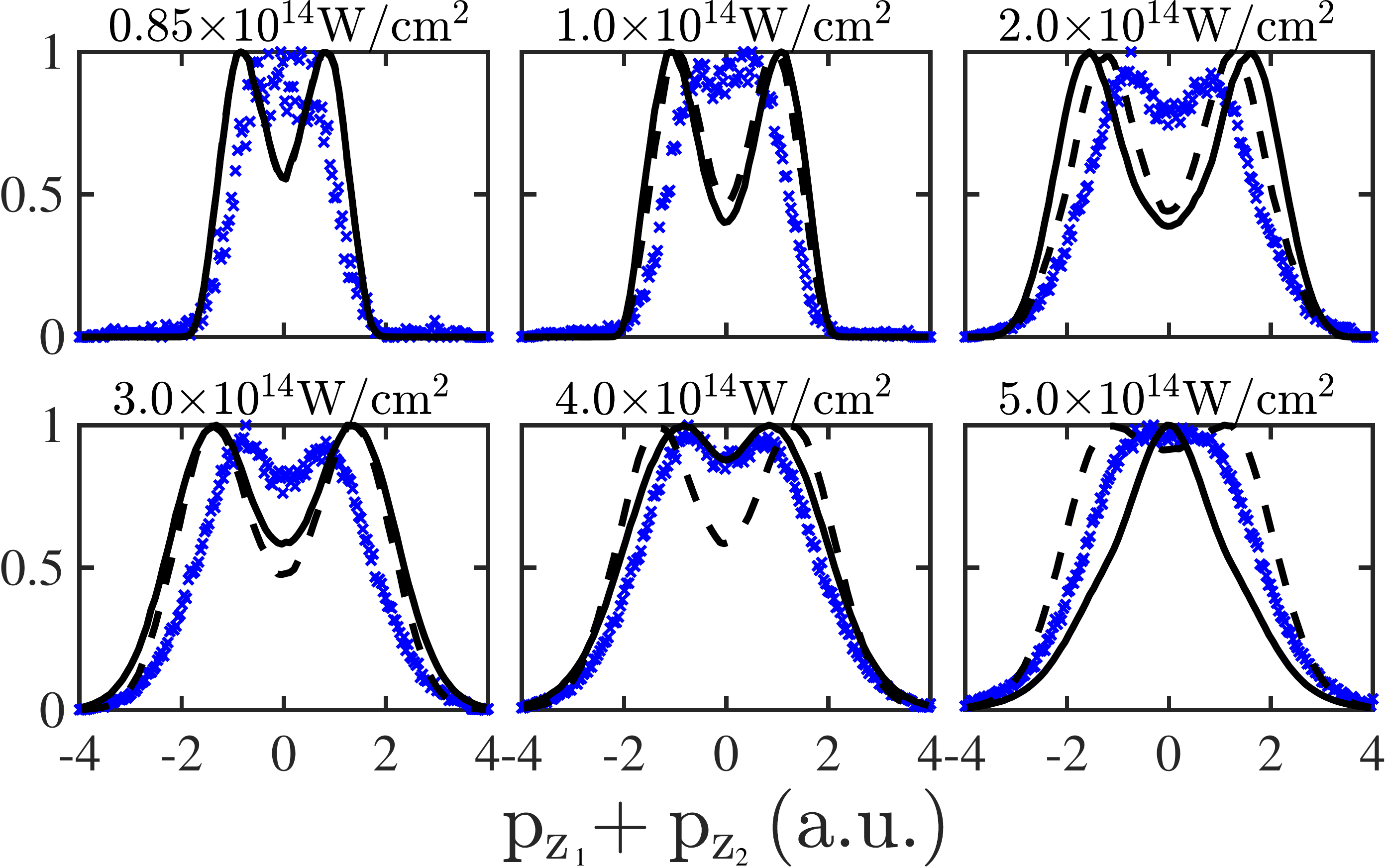}
\caption{Distribution of the sum of the two electron momentum components parallel to the polarization of the laser field for laser field intensities from 0.85$\times$10$^{14}$ W/cm$^{2}$ to 5$\times$10$^{14}$W/cm$^2$. The computed results with the focal volume effect not accounted for are denoted by black solid lines and when it is accounted for by black dashed lines; the blue crosses denote the experimental results.  }
\label{summomenta2}
\end{figure}

 In \fig{summomenta1}, the sum of the two electron momentum components along the polarization direction of the laser field  are presented for intensities from 0.85$\times$10$^{14}$ W/cm$^{2}$ to 5$\times$10$^{14}$W/cm$^2$. In \fig{summomenta1},   the contribution to the sum of the momenta of the direct and the delayed pathways of double ionization is also shown; the focal volume effect is not accounted for. As expected, it is found that the delayed pathway's contribution is a distribution concentrated  around zero while the direct pathway's contribution is a doubly-peaked distribution. The direct pathway's distribution of the sum of the momenta is the broadest one.   Therefore, including only the delayed pathway of double ionization  would result in a narrower distribution of the sum of the momenta than the  observed one. Indeed, the 1D model described in ref.\cite{Bergues} which accounts  only for the delayed  pathway of double ionization   results in a narrower distribution of the sum of the momenta than the observed one.  
In \fig{summomenta2},  the experimental results for the sum of the momenta in ref.\cite{Bergues} are compared with one set of computed results that account for the focal volume effect (black dashed lines) and one that does not (black solid lines). It is found that the computed results  are in good agreement with the observed ones. Specifically, it is found that, for each intensity,  the computed sum of the momenta extends over a range that  is very similar to the experimental one. For instance, for an intensity of 0.85$\times$10$^{14}$ W/cm$^{2}$,  the computed sum of the momenta extends over a range from roughly -2 a.u. to 2 a.u., while, for an intensity of 5$\times$10$^{14}$ W/cm$^{2}$, it  extends from -4 a.u. to 4 a.u.; for both intensities these ranges are in agreement with the experimental results \cite{Bergues}. It is noted that a difference of the computed distributions of the sum of the momenta with the experimental ones is that  the computed ones have smaller values around zero. This is more so the case for the computed results that account for the focal volume effect. This difference suggests that the current 3D model underestimates the contribution of the delayed pathway of double ionization.

 \subsection{Asymmetry parameter}
\begin{figure} [ht!]
\centering
 \includegraphics[clip,width=0.8\textwidth]{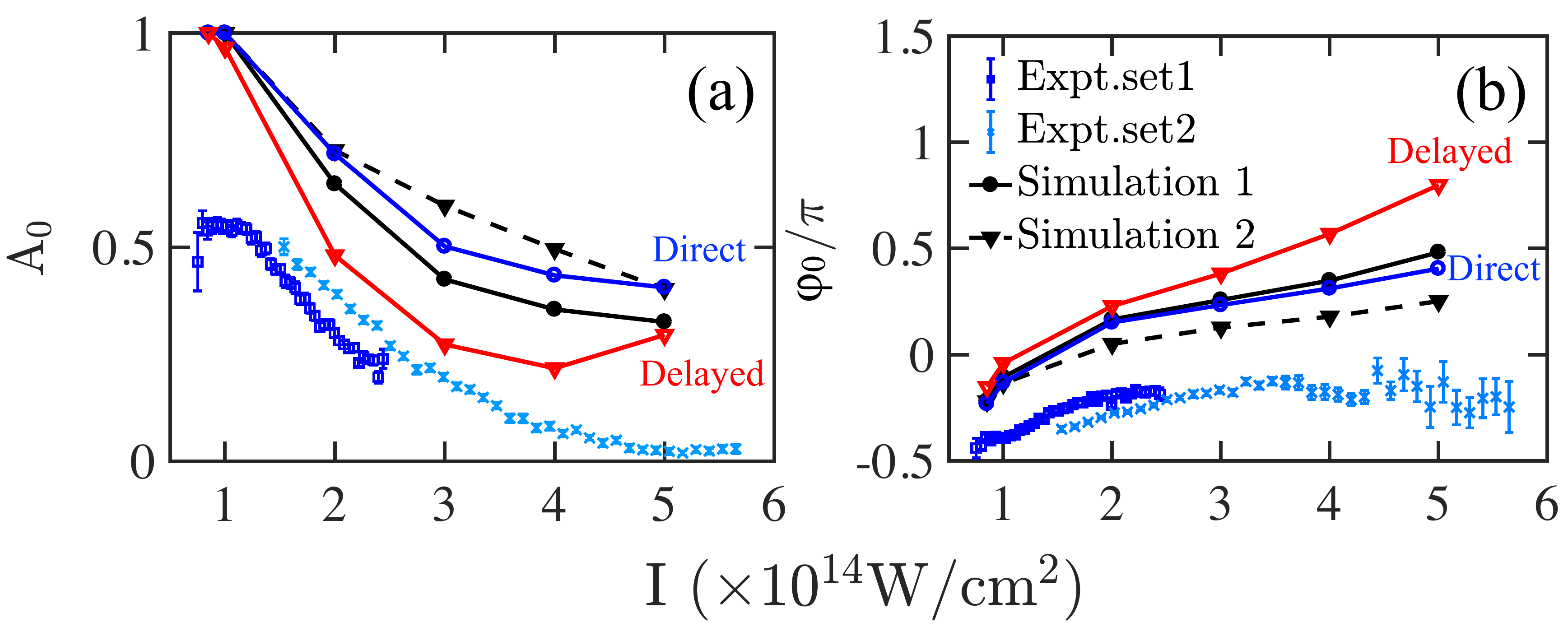}
\caption{Asymmetry parameters $A_{0}$ (a) and $\phi_{0}$ (b) as a function of intensity. The computed results for all double ionization events when the focal volume effect is not accounted for are denoted by black solid lines  with circles and when it is accounted for by  black dashed lines  with triangles. The delayed pathway of double ionization is denoted by red solid lines with triangles and   the direct pathway  by blue solid lines with circles. For the direct and delayed pathways of double ionization the focal volume effect is not accounted for. Experimental results are denoted by light blue crosses and dark blue squares.   }
\label{A0as}
\end{figure}

The asymmetry parameter 
\begin{equation}
A(I,\phi)=\frac{N_{+}(I,\phi)-N_{-}(I,\phi)}{N_{+}(I,\phi)+N_{-}(I,\phi)}
\label{eqasym1}
\end{equation}
is computed as a function of the intensity I and the CEP ($\phi$). $N_{+}(I,\phi)$ and $N_{-}(I,\phi)$ denote the percentage of double ionization events with ions escaping with positive and negative 
momentum, respectively, along the direction of the polarization of the laser field. Since in the 3D semiclassical model the nucleus is fixed,  $N_{+}(I,\phi)$ and $N_{-}(I,\phi)$ correspond to the percentage of double ionization 
events where the sum of the two electrons momentum components along the direction of the laser field polarization are negative and positive, respectively. 
For each intensity, $A(I,\phi)$ is fitted with the sinusoidal function 
\begin{equation}
A(I,\phi)=A_{0}(I)\sin{(\phi+\phi_{0}(I))}.
\end{equation}
The resulting asymmetry amplitude  $A_{0}(I)$ and offset phase $\phi_{0}(I)$ are plotted in \fig{A0as} (a) and (b), respectively, and  compared with  two sets of  experimentally obtained asymmetry parameters \cite{Bergues}. The comparison shows that the 3D semiclassical model reproduces well the  decreasing pattern of $A_{0}$ and the increasing pattern  of $\phi_{0}$ with increasing intensity. However, the computed values for these asymmetry parameters are higher than the ones obtained from the experimental results. In addition, in \fig{A0as} (a) and (b) the asymmetry parameters are plotted for each of the main two  pathways of NSDI. It is shown that for both pathways the asymmetry parameter $\phi_{0}(I)$ has a similar pattern.  The asymmetry parameter $A_{0}(I)$ for the delayed pathway is generally smaller which suggests that for the delayed pathway the two electron momentum components are more spread out in all four quadrants than for the direct ionization pathway. This is indeed shown in the correlated momenta presented in \fig{correlated} in the next section.

\subsection{Correlated momenta as a function of intensity: transition from strong to soft recollisions}
 
 \begin{figure} [ht!]
\centering
 \includegraphics[clip,width=0.90\textwidth]{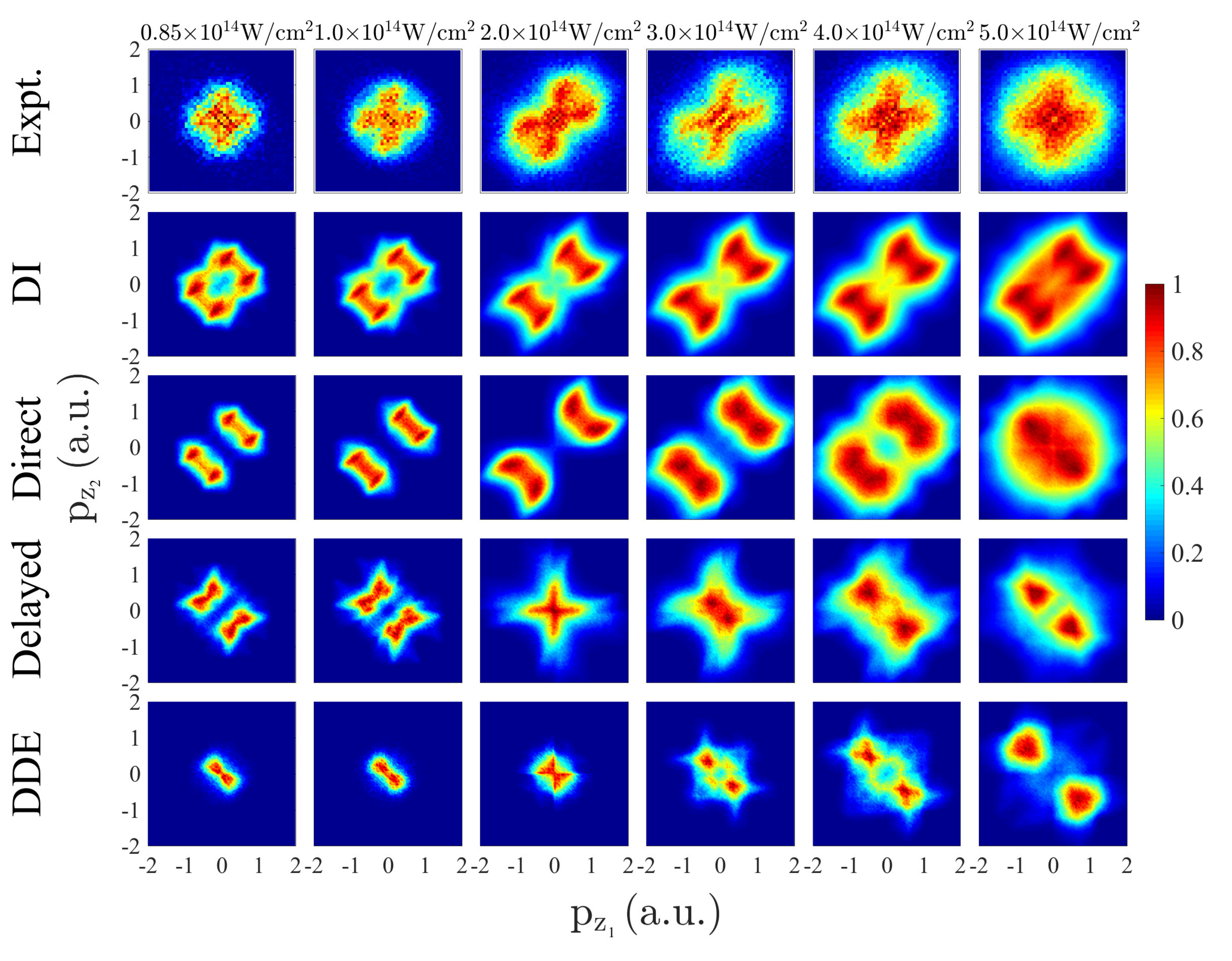}
\caption{First row: measured correlated momenta.
Second row: computed correlated momenta for all double ionization events with the focal volume effect accounted for. Third row: correlated momenta for the direct pathway of double ionization  with the focal volume effect not accounted for.
 Fourth row: correlated momenta for the delayed pathway of double ionization with the volume effect not accounted for. Fifth row: correlated momenta for the double delayed pathway of double ionization with the volume effect not accounted for.}
\label{correlated}
\end{figure}

 \begin{figure} [ht!]
\centering
 \includegraphics[clip,width=0.450\textwidth]{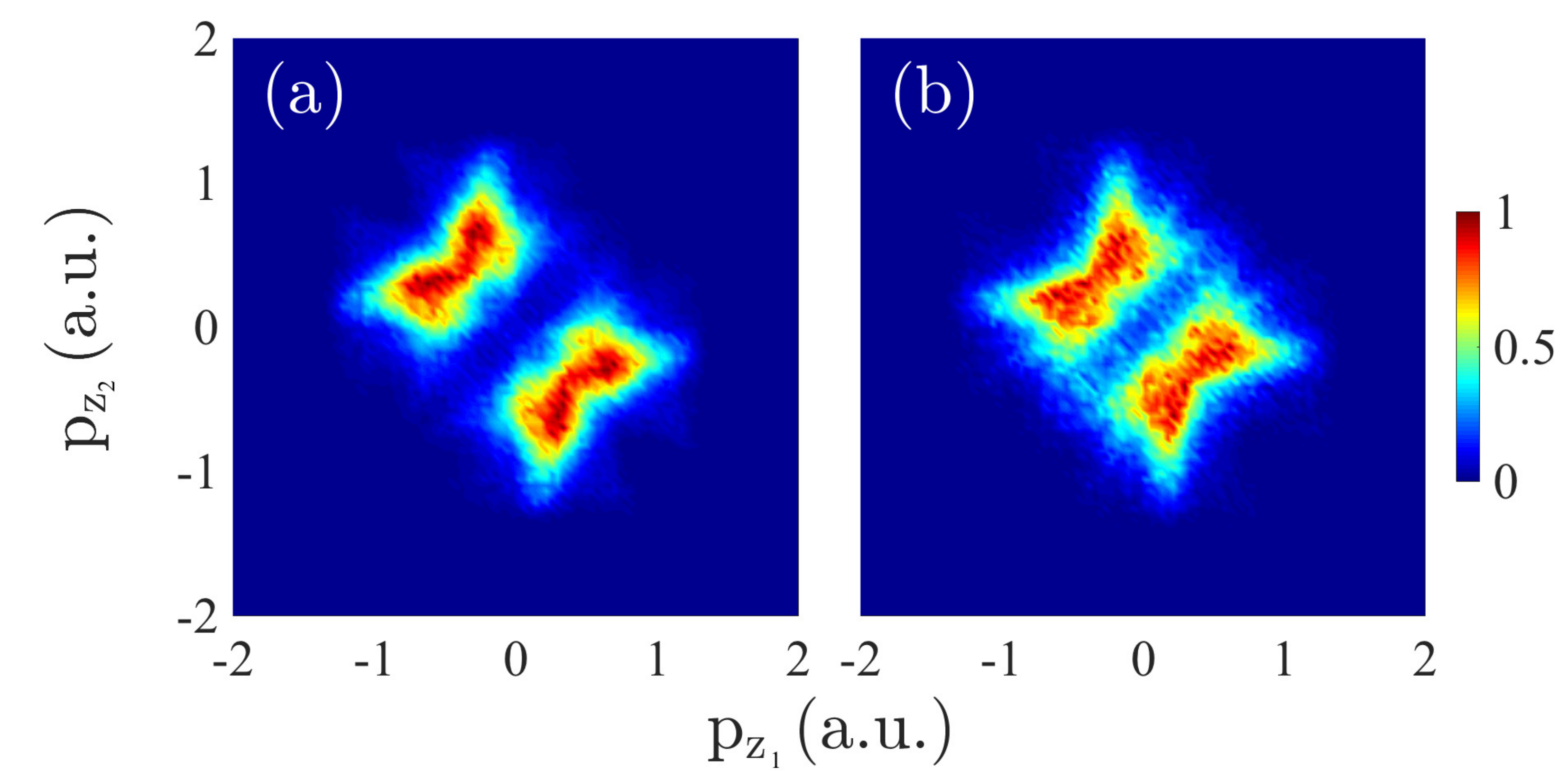}
\caption{Correlated momenta at an intensity 0.85$\times$10$^{14}$ W/cm$^{2}$  for the delayed pathway of double ionization for the case when the electron that ionizes second is electron 2 (a) and electron 1 (b).}
\label{correlated3}
\end{figure}

 For intensities ranging from 0.85$\times$10$^{14}$ W/cm$^{2}$ to 5$\times$10$^{14}$W/cm$^2$, the correlated momenta for the direct and delayed double ionization pathways are   plotted in  \fig{correlated}. For the smaller intensity of 0.85$\times$10$^{14}$ W/cm$^{2}$ the computed correlated momenta resemble but do not quite have the cross-shape pattern of the measured results, see \fig{correlated}. There are fewer double ionization events with both momenta being close to zero than in the experimentally obtained cross-shaped pattern \cite{Kling1, Bergues}. These close-to-zero momenta also reduce the asymmetry parameter. Thus, the reduced number of such events produced in our calculation is in accord with computing an asymmetry parameter $A_{0}(I)$  that is larger than the experimental one, see \fig{A0as}.
 The cross-shaped pattern is better reproduced  by the delayed double ionization pathway, see  \fig{correlated}. In addition, from an analysis of the classical trajectories it is found that for 63\% of the delayed pathway of double ionization  the initially bound electron ionizes after the tunneling electron  following recollision while for 37\% it is the other way around. The electron that ionizes second does so generally with a smaller momentum.  It is found that the correlated momenta  when the tunneling electron ionizes second resembles more a cross-shaped pattern, see \fig{correlated3}. 
 Moreover, as in the observed correlated momenta in ref.\cite{Bergues}, the computed correlated momenta transition  to the well-known pattern for intermediate intensities \cite{Weber}. This pattern involves both electrons escaping in the same direction either parallel or antiparallel to the laser field, thus, giving rise to a much higher probability density in the first and third quadrants of the correlated momenta, rather than the second and fourth ones. This is the pattern of the correlated momenta  for the direct pathway of  double ionization at intensities 2-4$\times$10$^{14}$ W/cm$^{2}$. At these intensities the direct pathway of double ionization  is the prevailing one.
 \begin{figure} [ht!]
\centering
 \includegraphics[clip,width=0.80\textwidth]{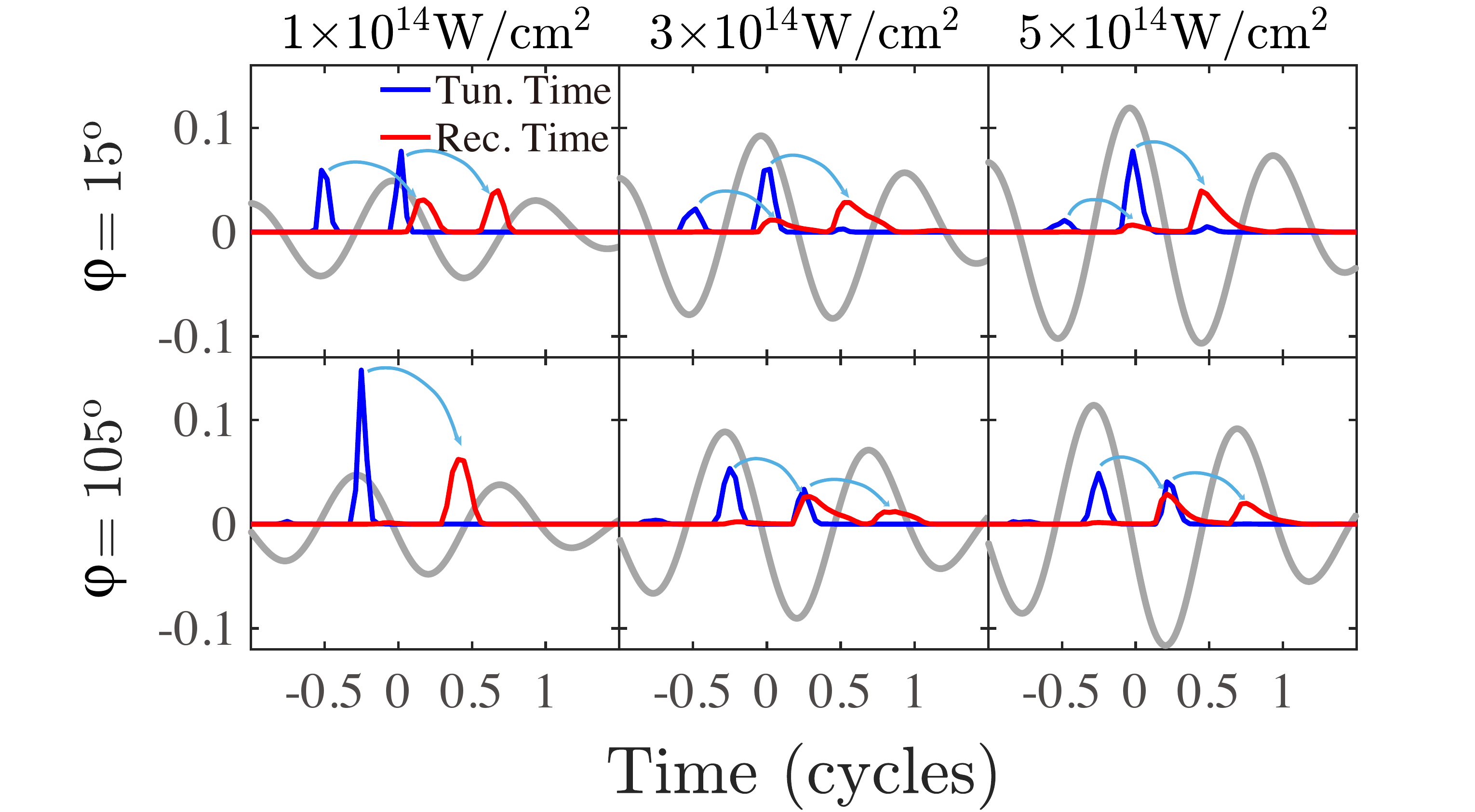}
\caption{Distribution of the tunneling time of electron 1 (blue line) and of the recollision time (red line) for intensities 10$^{14}$ W/cm$^{2}$, 3$\times$10$^{14}$ W/cm$^{2}$ and 5$\times$10$^{14}$ W/cm$^{2}$ and for two different CEPs, $\phi=15^{\circ}$ and $\phi=105^{\circ}$, for each intensity. Similar results hold for all other CEPs. Grey line denotes the laser field. The light blue arrows indicate the mapping of a tunneling-time peak to a recollision-time peak.}
\label{colltimes}
\end{figure}

  A less known pattern is that observed experimentally and retrieved computationally with the 3D semiclassical model for intensities above 4$\times$10$^{14}$ W/cm$^{2}$, see \fig{correlated}.  For these higher intensities, it is found that  the two electrons escape mostly with opposite momenta for a significant number of double ionization events. To identify the reason for this shift,  in \fig{colltimes},
the time electron 1 tunnel-ionizes and the recollision time are plotted for three different intensities, namely,  10$^{14}$ W/cm$^{2}$, 3$\times$10$^{14}$ W/cm$^{2}$ and 5$\times$10$^{14}$ W/cm$^{2}$ and for two different CEP's, namely, $\phi=15^{\circ}$ and $\phi=105^{\circ}$ for each intensity.  The tunneling time of electron 1 is found to be close to the times corresponding to the extrema of the laser field for all three intensities. However, the distribution of the  recollision time is found to shift from times corresponding  roughly to zeros of the laser field for an intensity of 10$^{14}$ W/cm$^{2}$ to  times corresponding to the extrema of the laser field for an intensity of 5$\times$10$^{14}$ W/cm$^{2}$. The transfer of energy from electron 1 to electron 2 is much smaller for the soft recollisions. For these higher intensities, where soft recollisions prevail, the momentum of electron 1 is mostly determined from the vector potential at the tunneling time. The momentum of electron 2 is determined by the vector potential shortly after recollision takes place which is roughly half a laser cycle after electron 1 tunnel-ionizes. As a result the two electrons escape mostly with opposite momenta. This mechanism of soft recollisions for higher intensities was first  identified in  a theoretical study of strongly-driven N$_{2}$ with fixed nuclei \cite{Agapi4}.  

The contribution to the correlated momenta of the direct and delayed pathways of double ionization is shown in \fig{correlated}. For the direct pathway, for intensities from 0.85$\times$10$^{14}$ W/cm$^{2}$ to 4$\times$10$^{14}$W/cm$^2$ the probability density is significantly higher in the first and third quadrants of the correlated momenta than in the second and fourth ones. This is a known  pattern. For strong recollisions, the two electron momentum components along the direction of the laser field 
are both determined from the vector potential at times just larger than the recollision  time. Thus, both electrons escape with similar momenta in the direction along the polarization of the laser field. For an intensity of 5$\times$10$^{14}$ W/cm$^{2}$ soft recollisions 
prevail and, as discussed above,  the two electrons escape mostly with opposite  momenta along the direction of the polarization of the laser field. For the delayed pathway, this opposite momenta pattern, demonstrated  with much higher probability density in the second and fourth quadrants of the correlated momenta, sets in at lower intensities, 3$\times$10$^{14}$ W/cm$^{2}$, see  \fig{correlated}. This is consistent with a smaller transfer of energy taking place from electron 1 to electron 2 at the recollision time in the delayed pathway compared to the energy transfer in the direct pathway. Regarding small intensities, in \fig{correlated} it is shown  that the computed correlated momenta due to the delayed pathway have a pattern similar to but not quite cross-shaped.

\subsection{Correlated momenta and double ionization pathways as a function of CEP}
\begin{figure} [ht!]
\centering
 \includegraphics[clip,width=0.7\textwidth]{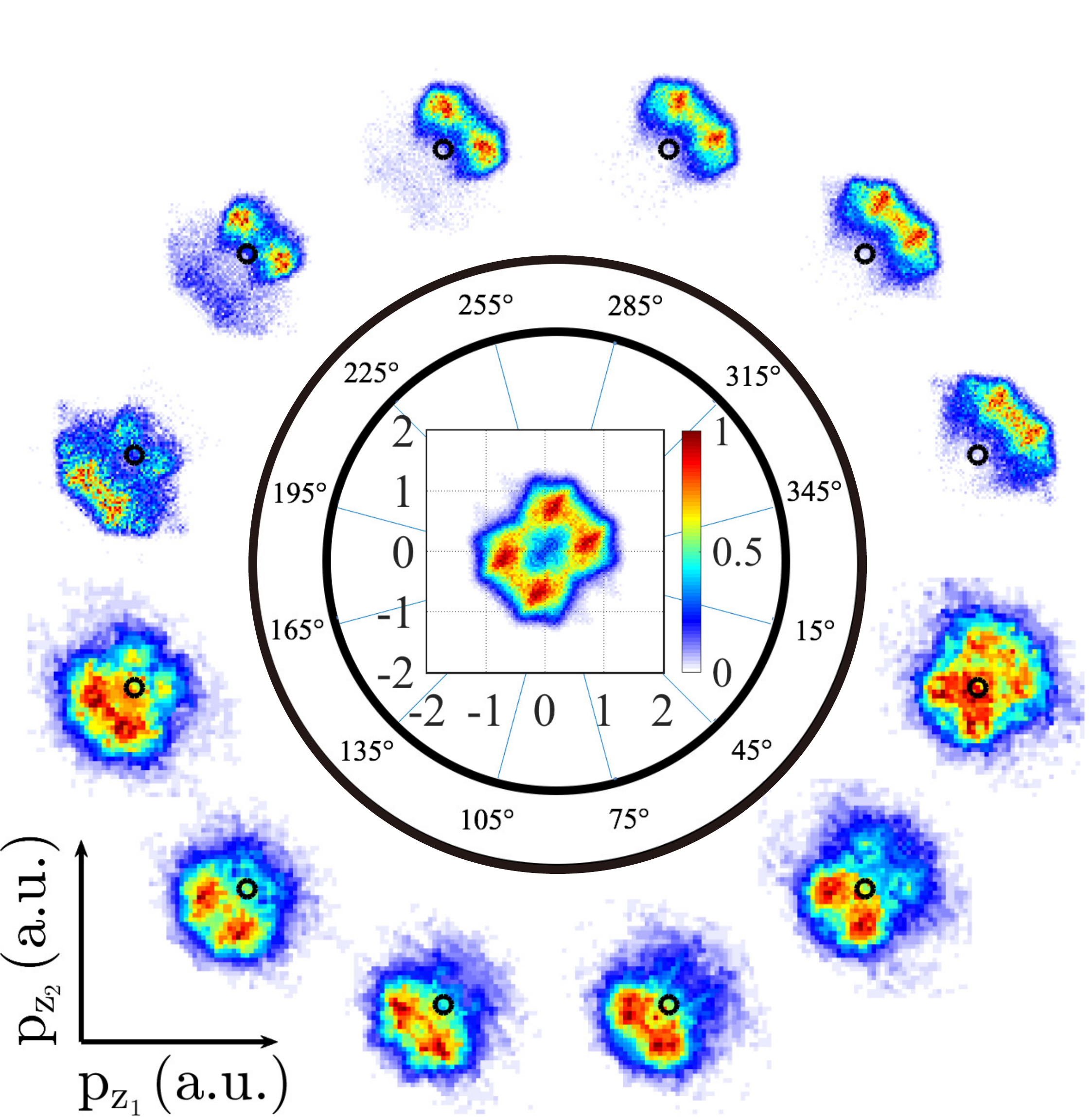}
\caption{Correlated momenta at an intensity of 0.85$\times$10$^{14}$ W/cm$^{2}$ for   $\phi$ ranging from $15^{\circ}$ to $345^{\circ}$ with a step of $30^{\circ}$. For  $\phi$ ranging from $15^{\circ}$ to $165^{\circ}$ the experimental results from  ref.\cite{Kling3} are displayed. For  $\phi$ ranging from $195^{\circ}$ to $345^{\circ}$ the computed results are displayed. The average over all $\phi$s using the computed results is the correlated momentum plot in the middle. The dark circle in each plot denotes the origin of the axes.}
\label{CEPcor}
\end{figure}

\begin{figure} [ht!]
\centering
 \includegraphics[clip,width=0.40\textwidth]{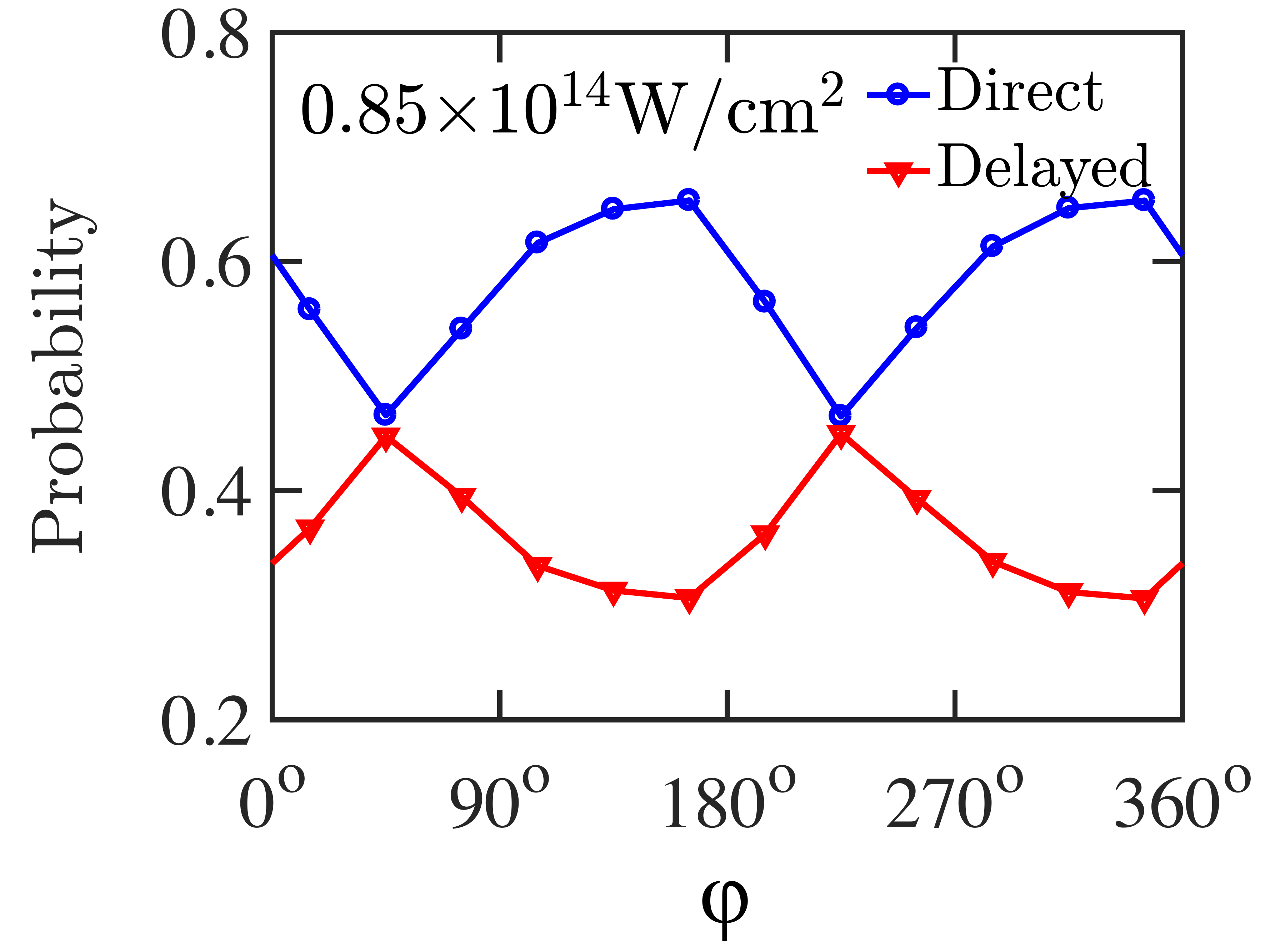}
\caption{\% of the direct and the delayed pathways of double ionization events as a function of the CEP for an intensity of  0.85$\times$10$^{14}$ W/cm$^{2}$. }
\label{CEPprob}
\end{figure}

In what follows,  the dependence of the correlated momenta on the CEP is investigated at an intensity of 0.85$\times$10$^{14}$ W/cm$^{2}$. In \fig{CEPcor}, the correlated momenta are plotted for $\phi$ ranging from $15^{\circ}$ to $345^{\circ}$ with a step of $30^{\circ}$. Due to the symmetry of the Hamiltonian, when  $\phi \rightarrow \phi+180^{\circ}$  then ${\bf p} \rightarrow {\bf -p}$. With this symmetry in mind,  the experimental results from ref.\cite{Kling3} are plotted in \fig{CEPcor} for $\phi$ ranging from $15^{\circ}$ to $165^{\circ}$  and the computed results are plotted for $\phi$ ranging from $195^{\circ}$ to $345^{\circ}$.   A very good agreement is found between the computed and the experimental results. Specifically, the computed correlated momenta correctly reproduce the overall observed pattern for each individual CEP.  
A difference between the computed and the experimental results is that the former results are more concentrated on the first and third quadrants suggesting that the computations overestimate the contribution of the direct pathway. To better understand the change of the correlated momenta pattern as a function of the CEP plotted in \fig{CEPcor}, in \fig{CEPprob} the percentage of the direct and delayed pathways of double ionization events are plotted as a function of the CEP. At an intensity of 0.85$\times$10$^{14}$ W/cm$^{2}$ the delayed pathway has the largest contribution for CEPs $\phi=45^{\circ}$ and 
$\phi=225^{\circ}$, while it has the smallest for CEPs $\phi=165^{\circ}$ and $\phi=345^{\circ}$. In \fig{CEPcor},  a comparison of the correlated momenta between $\phi=45^{\circ}$ and $\phi=165^{\circ}$ 
shows that there is higher probability density for both electrons to ionize with the same large momentum for $\phi=165^{\circ}$ than for $\phi=45^{\circ}$. This is indeed consistent with the direct ionization pathway having a larger contribution 
for $\phi=165^{\circ}$ than for $\phi=45^{\circ}$ as shown in \fig{CEPprob}. Similar conclusions can be drawn by comparing the correlated momenta for $\phi=345^{\circ}$ with  the one for  $\phi=225^{\circ}$. From \fig{CEPprob}, it is  found that  the 
contribution of each of the two main pathways of double ionization vary roughly by 20\% as a function of the CEP for the smallest intensity  of 0.85$\times$10$^{14}$ W/cm$^{2}$. 

\section{Conclusions}
Using a 3D semiclassical model we investigated the dependence of double ionization observables on the intensity and on the carrier envelope phase of a near-single-cycle near-infrared laser field employed to drive Ar. The good agreement of the computed results  with recent experiments  employing near-single-cycle laser pulses \cite{Kling1,Kling3, Bergues}, adds to previous successes of this 3D model  in identifying features of non-sequential double ionization of two-electron atoms when driven by many-cycle laser pulses \cite{Agapi1,Agapi2, Emmanouilidou1}. A difference between the computed and  the experimental results was found to be a lower value of the distribution of the sum of the two electron  momentum components along the direction of the polarization 
of the laser field and a lower value around zero of the correlated momenta. This seems to suggest that the current 3D model overestimates the Coulomb attraction of each electron from the nucleus. Future studies can improve on the 3D model for many electron atoms such as Ar by using more accurate effective potentials for the time propagation.  Moreover, it was demonstrated that the main pathways of double ionization, that is, the direct and the delayed pathways, both significantly contribute at all intensities currently under consideration. Furthermore, the prevalence of the direct versus the delayed pathway was investigated as a function of the CEP for an intensity of  0.85$\times$10$^{14}$ W/cm$^{2}$ and it was shown that the results obtained are consistent with features of the observed correlated momenta \cite{Kling3}. Finally, a previously-predicted in the context of  a strongly-driven fixed-nuclei N$_{2}$ unexpected anti-correlation momentum pattern at higher intensities \cite{Agapi4}, is observed experimentally  in the context of strongly-driven Ar \cite{Bergues} and also reproduced in the current work for strongly-driven Ar by a near-single-cycle laser field. It is shown that this anti-correlation   pattern is due to soft recollisions with recollision times close to the extrema of the laser field.

%

%

\section*{Acknowledgements}
A.E.  acknowledges the EPSRC grant no. J0171831 and the use of the computational resources of Legion at UCL. M.K., M.F.K., and B.B. acknowledge support from the Max Planck Society and the DFG cluster of excellence ``Munich Centre for Advanced Photonics (MAP)".

\section*{Author contributions statement}

A.C. performed the analysis of the computations involved and contributed to the ideas involved. A.E.  provided the codes used for the computations and the analysis and was responsible for the main ideas involved in the theoretical analysis. M.K., M.F.K., and B.B provided the experimental results and contributed to the discussion in the paper concerning the comparison of the computations with  the experimental results. All authors reviewed the manuscript. 

\section*{Additional information}
\textbf{Competing financial interests:} The authors declare no competing financial interests.
%

\end{document}